\documentstyle[prd,aps]{revtex}
\newcommand{\be}{\begin{equation}}
\newcommand{\ee}{\end{equation}}
\newcommand{\bea}{\begin{eqnarray}}
\newcommand{\eea}{\end{eqnarray}}
\def\aprle{\buildrel < \over {_{\sim}}}
\def\aprge{\buildrel > \over {_{\sim}}}
\begin{document}
\draft
\input epsf
\twocolumn[\hsize\textwidth\columnwidth\hsize\csname
@twocolumnfalse\endcsname

\title{Linking solar and long baseline terrestrial 
neutrino experiments}
\author{E. Kh. Akhmedov${}^{(a,b)}$ 
G. C. Branco${}^{(a)}$ 
and M. N. Rebelo${}^{(a)}$ 
}
\address{$^{(a)}${\em
CFIF, Departamento de Fisica, Instituto Superior T\'ecnico,
P-1049-001 Lisboa, Portugal}}
\address{${}^{(b)}${\em 
National Research Centre Kurchatov Institute, Moscow 123182, Russia}} 
\date{December 1, 1999}
\maketitle
\begin{abstract}
We show that 
in the framework of three light neutrino species with hierarchical masses 
and assuming no fine tuning between the entries of the neutrino mass matrix, 
one can use the solar neutrino data to obtain information on the element 
$U_{e3}$ of the lepton mixing matrix. Conversely, a measurement of $U_{e3}$ 
in atmospheric or long baseline accelerator or reactor neutrino experiments 
would help discriminate between possible oscillation solutions of the solar 
neutrino problem. 
\end{abstract}
\pacs{PACS: 14.60.Pq, 13.15.+g  \hskip 2 cm FISIST/21-99/CFIF 
\hskip 2cm hep-ph/9912205}
\vskip1.2pc ]

\noindent
1. Currently, there are indications for neutrino oscillations in solar
\cite{SNP}, atmospheric \cite{ANA} and accelerator \cite{LSND} 
experiments, with the strongest evidence coming from the Super-Kamiokande 
atmospheric neutrino data \cite{ANA}. If all correct, these results would 
imply the existence of at least four light neutrino species, $\nu_e$, 
$\nu_\mu$, $\nu_\tau$ and $\nu_s$, where $\nu_s$ is a sterile (electroweak 
singlet) neutrino. Of the above mentioned experimental evidence, the 
result of the accelerator LSND experiment is the only one that has not yet
been independently confirmed. 
If it is excluded, the remaining solar and atmospheric neutrino 
anomalies can be explained through oscillations among just three 
standard neutrinos -- $\nu_e$, $\nu_\mu$ and $\nu_\tau$. The oscillation 
probabilities for relativistic neutrinos then depend on two mass squared 
differences $\Delta m_{21}^2 \equiv \Delta m_\odot^2$ and $\Delta m_{32}^2 
\equiv \Delta m_{atm}^2$, three mixing angles $\theta_{12}$, $\theta_{13}$ 
and $\theta_{23}$, and one CP-violating phase $\delta$. With the 
parametrization of the $3\times 3$ leptonic mixing matrix $U$ which coincides 
with the standard parametrization of the quark mixing matrix \cite{PDG}, 
one can identify the mixing angle which is responsible for the dominant 
channel of the atmospheric neutrino oscillations with $\theta_{23}$, the one 
that is primarily responsible for the solar neutrino oscillations with
$\theta_{12}$ and the mixing angle which enters (along with $\theta_{23}$)
into the probabilities of the subdominant $\nu_e\leftrightarrow 
\nu_{\mu(\tau)}$ oscillations of atmospheric neutrinos and long baseline 
$\nu_e\leftrightarrow \nu_{\mu(\tau)}$ oscillations with $\theta_{13}$. 
For the values of the neutrino parameters allowed by the data, the 
CP-violating effects in neutrino oscillations should be rather small, and 
we shall therefore omit the phase $\delta$ in our analysis. 

The Super-Kamiokande atmospheric neutrino data imply $\Delta m_{32}^2\simeq 
(2 - 6)\times 10^{-3}$ eV$^2$, $\theta_{23}\simeq (45 \pm 12)^\circ$, and the 
combined data of the solar neutrino experiments lead to four domains of 
allowed values of $\Delta m_{21}^2$ and $\theta_{12}$ corresponding to the
four neutrino oscillation solutions to the solar neutrino problem -- large
mixing angle MSW (LMA), small mixing angle MSW (SMA), vacuum 
oscillations (VO) and low-$\Delta m^2$ (LOW) solutions \cite{f1}. 
The LOW solution has a low probability and is often excluded from 
discussions. The remaining mixing angle $\theta_{13}$ which determines the 
element $U_{e3}$ of the lepton mixing matrix is the least known one: there 
are only upper limits on its value, the most stringent one coming from the 
CHOOZ reactor neutrino experiment \cite{CHOOZ}. 
Together with the solar neutrino observations it gives, 
for $\Delta m_{32}^2 =(2 - 6)\cdot 10^{-3}~{\rm eV}^2$, 
\be
|\sin \theta_{13}|\equiv |U_{e3}| \le (0.22 - 0.14)\, \quad
\label{chooz}
\ee
The probabilities of the long baseline $\nu_e\leftrightarrow \nu_{\mu(\tau)}$ 
oscillations and subdominant $\nu_e\leftrightarrow \nu_{\mu(\tau)}$ 
oscillations of atmospheric neutrinos depend sensitively on $\sin\theta_{13}$, 
and therefore knowledge of its value at least by an order of magnitude 
would be very helpful for planning future long baseline experiments.
Yet, the upper limit (\ref{chooz}) does not tell us what this value is -- 
it can equally well be just below the upper bound or many orders of 
magnitude smaller.  

In the present letter we show how one can extract information on the 
value of $U_{e3}$ from the solar neutrino data under the
assumption that there is no fine tuning between certain entries of
the neutrino mass matrix. 
We shall  derive predictions for $U_{e3}$ corresponding to each one of the
neutrino oscillation solutions to the solar neutrino problem. \\

\noindent
2. In the three-flavour framework, assuming the hierarchy $\Delta m_{21}^2 
\ll \Delta m_{32}^2$, the survival probability of the solar $\nu_e$ can 
be written as 
\cite{prob}
\be
P_S \simeq c_{13}^4 P + s_{13}^4\,,
\label{prob}  
\ee
where we use the notation $s_{ij}\equiv \sin\theta_{ij}$,  $c_{ij}\equiv
\cos\theta_{ij}$ and  $P$ is the corresponding survival probability in the 
two-flavour case which depends on the mixing angle $\theta_{12}$ and mass
squared difference $\Delta m_{21}^2$, with the usual matter-induced
potential for neutrinos $V=\sqrt{2}G_F N_e$ \cite{MSW} replaced by the 
effective one $V_{eff}=c_{13}^2 V$.  
It follows from (\ref{prob}) that $P_S$ is rather insensitive to the value of 
$\theta_{13}$ provided that the constraint (\ref{chooz}) is satisfied.  
Therefore the probability of the solar neutrino oscillations cannot be 
used directly to extract a useful information on $U_{e3}$. We shall show, 
however, that such an information can still be obtained from the  
analyses of the neutrino mass matrix provided that the values of 
the parameters that govern the solar neutrino oscillations are known. 

Assuming the neutrino mass hierarchy $m_1,m_2\ll m_3$ and $\theta_{23}\simeq 
45^\circ$ (which is the best fit value of the Super-Kamiokande data 
\cite{ANA}) and taking into account the relative smallness of 
$\theta_{13}$, it can be shown that 
in the basis where the mass matrix of charged leptons is diagonal 
the neutrino mass matrix $m_L$ must have the approximate form 
\be
m_L = m_0 \left(\begin{array}{ccc}
\kappa      & \varepsilon     & \varepsilon' \\
\varepsilon & ~1+\delta-\delta' & 1-\delta \\   
\varepsilon' & ~1-\delta & 1+\delta+\delta'
\end{array}
\right)\,,
\label{mL1}
\ee
where $\kappa$, $\varepsilon$, $\varepsilon'$, $\delta$ and $\delta'$ are 
small dimensionless parameters. Diagonalization of this matrix \cite{Akh} 
yields, in particular, 
\be
\tan 2\theta_{12} \simeq
\frac{(\varepsilon-\varepsilon')}
{\sqrt{2}\left[\left(\delta-\frac{\delta'^{2}}{4}\right)-\left(\frac{\kappa}
{2}-\frac{\varepsilon^2+\varepsilon'^{2}}{4}\right)\right]}\,,
\label{t2}
\ee
\be
s_{13} \simeq \frac{\varepsilon+\varepsilon'}{2\sqrt{2}}\,.
\label{s13a}
\ee
We shall be assuming that there are no accidental cancellations between 
$\varepsilon$ and $\pm\varepsilon'$, i.e. that $|\varepsilon+\varepsilon'|$ 
and $|\varepsilon-\varepsilon'|$ are of the same order of magnitude: 
$|\varepsilon \pm \varepsilon'| \sim \tilde{\varepsilon}$ where 
$\tilde{\varepsilon}=max\{|\varepsilon|, |\varepsilon'|\}$. 
In other words, we assume that $|\varepsilon|$ and $|\varepsilon'|$ are
either of the same order of magnitude or one of them is much larger than the 
other, but bar the possibility that they are equal or approximately 
equal to each other. 

It has been shown in \cite{Akh} that the MSW effect \cite{MSW} can only
occur for neutrinos, and in particular the LMA and SMA solutions of the 
solar neutrino problem are only possible, if the parameters of the mass 
matrix $m_L$ in Eq. (\ref{mL1}) satisfy
\be
|\tilde{\delta}|\equiv |\delta-\delta'^{2}/4| >
|{\kappa}/2-(\varepsilon^2+\varepsilon'^{2})/4|\,.
\label{cond}
\ee
We shall first assume $|\tilde{\delta}| \gg \varepsilon^2, \varepsilon'^{2}, 
|\kappa|$  (our results will also be approximately valid when $\gg$ is 
replaced by $\aprge$). From (\ref{t2}) and (\ref{s13a}) one finds 
\be
\tan 2\theta_{12}\simeq 
\tilde{\varepsilon}/\sqrt{2}\,\tilde{\delta}\,, \quad 
s_{13}\simeq \tilde{\varepsilon}/2\sqrt{2}\,. 
\label{thetaepsilon}
\ee
The eigenvalues of the mass matrix $m_L$ can then approximately be written as 
\be
m_{1,2}\simeq m_0\,\tilde{\delta}\left( 1\pm\sqrt{1+\tan^2 2\theta_{12}}
\right)\,,\quad m_3 \simeq 2 m_0\,,
\label{m12b}
\ee
leading, with the identification $\Delta m_\odot^2=\Delta m_{21}^2$, 
$\Delta m_{atm}^2\simeq \Delta m_{32}^2\simeq (2m_0)^2$, to 
\be
\tilde{\delta} \simeq \frac{1}{(1+\tan^2 2\theta_{12})^{1/4}}\left(
\frac{\Delta m_\odot^2}{\Delta m_{atm}^2}\right)^{1/2}\,.
\label{delta1} 
\ee
{}From Eq. (\ref{thetaepsilon}) one then finds 
\be
s_{13}\simeq\frac{1}{2}\,\frac{\tan 2\theta_{12}}{(1+\tan^2
2\theta_{12})^{1/4}}\left(\frac{\Delta 
m_\odot^2}{\Delta m_{atm}^2}\right)^{1/2}\,.
\label{s13b} 
\ee
This expression gives, up to a factor of the order one, the value of the 
lepton mixing parameter $U_{e3}=s_{13}$ in terms of the parameters describing 
the solar neutrino  oscillations. 
Substituting the typical values of the parameters that lead to the 
various neutrino oscillation solutions of the solar neutrino problem
\cite{f1} we find 
\begin{eqnarray}
s_{13} & \simeq & (0.05 - 0.15) ~~{\rm (LMA)}\,; \quad   
\sim 10^{-3} ~~{\rm(SMA)}\,;
\nonumber \\
&\sim & 10^{-2}~~{\rm (LOW)}\,;
\quad \;\, \sim 10^{-4} - 10^{-3} ~~{\rm (VO)}\,.
\label{res1}
\end{eqnarray} 
Thus, in the case of the LMA solution the value of $s_{13}$ is expected 
to be only slightly below the CHOOZ limit. The values of $s_{13}$ in this 
range can lead to observable effects in the $\nu_e\leftrightarrow 
\nu_{\mu(\tau)}$ channels of the long baseline experiments as well as in the 
subdominant $\nu_e\leftrightarrow \nu_{\mu(\tau)}$ channels of the atmospheric 
neutrino experiments. 
They should certainly be detectable in MINOS \cite{minos} 
except perhaps for the values of $s_{13}$ close to the lower border of the
allowed region. However, in this case they should still be detectable in the 
future long baseline experiments with muon storage rings which are being 
widely discussed now \cite{Barger}. They may also be detectable in 
KamLAND \cite{kamland} and CERN -- Gran Sasso \cite{NGS} experiments 
provided that the value of $s_{13}$ is close to the upper border of the 
allowed region for the LMA solution. 

In the case of the LOW solution, the values of $s_{13}$ are 
close to the border of detectability in the experiments with muon storage
rings. Whether or not they will be detectable depends on the experimental 
details which are not yet known. For the SMA and VO solutions of the solar 
neutrino problem, the predicted values of $s_{13}$ are far too small to lead 
to observable effects in any of the forthcoming or currently discussed
long baseline experiments. 

Eq. (\ref{s13b}) is not valid when $\theta_{12}$ is very close to $45^\circ$, 
namely when $1-\sin^2 2\theta_{12}\aprle 10^{-5}$. Such a situation can in
principle be realized in the case of the VO and LOW solutions of the solar
neutrino problem \cite{f1,low}. In this case from (\ref{m12b}) and 
(\ref{thetaepsilon}) 
one finds $\Delta m_\odot^2 \simeq \tilde{\delta}^2 \tan 2\theta_{12}
\Delta m_{atm}^2 \simeq (\tilde{\delta} \tilde{\varepsilon}/\sqrt{2})
\Delta m_{atm}^2$. Our condition $|\tilde{\delta}|\aprge 
\tilde{\varepsilon}^2$ and Eq. (\ref{thetaepsilon}) then lead to the 
following upper limit on $s_{13}$: 
\be
(s_{13})_{max}\simeq 2^{-4/3}\,\left(\frac{\Delta m_\odot^2}{\Delta
m_{atm}^2}\right)^{1/3}\,.
\label{s13c} 
\ee
Consider now the case $|\tilde{\delta}|\ll \varepsilon^2, \varepsilon'^{2}, 
|\kappa|$, i.e. $|\tilde{\delta}|\ll \tilde{\varepsilon}^2$. Condition 
(\ref{cond}) is then not satisfied and therefore only the VO and LOW
solutions to the solar neutrino problem are possible. It is easy to show that 
in this case $s_{13}$ is also approximately given by Eq. (\ref{s13c}) which,
however, is now the prediction rather than an upper bound. For typical values 
of $\Delta m_\odot^2$ relevant for the VO and LOW solutions one then finds 
$s_{13}\sim 10^{-3}$ and $s_{13}\sim 10^{-2}$ respectively, which are in
the same ranges as the values given for these solutions in (\ref{res1}). \\

\noindent
3. The above discussion applied to the case of the normal neutrino mass 
hierarchy, $|m_{1,2}| \ll |m_3|$. Consider now the case of the inverted
mass hierarchy with $|m_3| \ll |m_1| \simeq |m_2|$. There are essentially 
two possibilities. First, the neutrino mass matrix can have the elements 
$(m_L)_{12}=(m_L)_{21} \simeq (m_L)_{13}=(m_L)_{31} = m_0$ with the rest
of the matrix elements being $\sim 10^{-8} m_0$. Such a matrix 
can emerge due to an approximate $L_e-L_\mu-L_\tau$ symmetry \cite{Barb}. 
It leads to the VO solution of the solar neutrino problem with bi-maximal 
mixing and $m_1\simeq -m_2$ (i.e. opposite CP-parities of the mass eigenstates
$\nu_1$ and $\nu_2$). The mixing parameter $s_{13}$ is given by the ratio
of a combination of the small entries of the mass matrix and $m_0$ \cite{Akh}, 
i.e. in this case 
\be
s_{13}\sim 10^{-8}\,,
\label{s13d}
\ee
far too small to be of any practical interest. 
Second, the neutrino mass matrix may again be of the
form (\ref{mL1}) with small parameters $\varepsilon$, $\varepsilon'$, 
$\delta$ and $\delta'$ but now with $\kappa \simeq \pm 2$, which is 
necessary for the eigenvalues of $m_L$ to satisfy $m_1 \simeq \pm m_2$. 
As before, one can express $s_{13}$ through the parameters describing the 
solar neutrino oscillations. Consider first the case $\kappa \simeq 2$, which 
leads to same sign $m_1$ and $m_2$ (same CP-parities of $\nu_1$ and 
$\nu_2$). In this case any of the neutrino oscillation solutions to the
solar neutrino problem can be accommodated. Using the results of \cite{Akh} 
one obtains, again up to a factor of the order one, 
\be
s_{13}\simeq\frac{1}{4}\,\sin 2\theta_{12}\left(\frac{\Delta 
m_\odot^2}{\Delta m_{atm}^2}\right)\,.
\label{s13e} 
\ee
The predicted numerical values of $s_{13}$ for various solutions of the 
solar neutrino problem are
\begin{eqnarray}
s_{13} & \simeq & (0.15 - 1.5)\times 10^{-2} ~~{\rm (LMA)}\,;
\;\; \sim 3\times 10^{-5} ~~{\rm(SMA)}\,;
\nonumber \\
&\sim & 10^{-5}~~{\rm (LOW)}\,;
\quad  \sim 10^{-8} ~~{\rm (VO)}\,,
\label{res2}
\end{eqnarray}  
too small to be of interest except perhaps for the LMA case which might 
lead to observable effects in future experiments with muon storage rings. 
  
Consider now the case $\kappa \simeq -2$, which 
leads to $m_1\simeq -m_2$ (opposite CP-parities of $\nu_1$ and $\nu_2$). In 
this case only the SMA solution to the solar neutrino problem can be 
accommodated \cite{f2}. Diagonalization of the mass matrix yields 
\be
s_{13}\simeq \tan 2\theta_{12}\,.
\label{s13f} 
\ee
Since the SMA solution requires $\sin^2 2\theta_{12}\simeq (0.1 - 1)\times
10^{-2}$, this gives 
\be
s_{13} \simeq 0.03 - 0.1\,,
\label{res3} 
\ee
i.e. one can have observable $\nu_e\leftrightarrow \nu_{\mu(\tau)}$ 
oscillations in the long baseline experiments in this case. \\

\noindent
4. The results we have obtained rely crucially on the assumption of no
fine tuning between certain elements of the neutrino mass matrix. Although 
we believe that this assumption is natural, such fine tuning is still a 
possibility; therefore our results should only be considered as the likely
values of the parameter $U_{e3}$. 

Eqs. (\ref{s13b}), (\ref{s13c}) -- (\ref{s13e}) and (\ref{s13f}) are our 
main results. They give, 
for various neutrino mass hierarchies and relative CP-parities, the 
approximate values of the lepton mixing parameter $U_{e3}$ in terms of the 
values of the parameters governing the oscillations of solar neutrinos. 
We have checked these relations by direct numerical diagonalization of the
neutrino mass matrix $m_L$ for a number of the parameter sets leading to the 
SMA, LMA, LOW and VO solutions of the solar neutrino problem and found
that in most of the cases the agreement was better than $50\%$.  

Our predictions for $U_{e3}$ depend on the assumed hierarchy of neutrino
masses. The normal mass hierarchy  $|m_{1,2}|\ll |m_3|$ is the most natural 
one; the mass matrices leading to the inverted mass hierarchy are unstable
with respect to small variations of the parameters except in the 
case when the elements $(m_L)_{12}=(m_L)_{21}$ and $(m_L)_{13}=(m_L)_{31}$ 
are much larger than the rest of the matrix elements \cite{Akh}. 
However, this case leads to the VO solution of the solar neutrino problem 
with an extremely small $U_{e3}$ of Eq. (\ref{s13d}). It should be noted that 
the question of the neutrino mass hierarchy can in principle be settled 
experimentally: the long baseline experiments may discriminate between the
direct and inverted hierarchies through the earth matter effects on
neutrino oscillations. 

If the LMA MSW effect proves to be the true solution of the solar neutrino 
problem, one can expect observable effects in the $\nu_e \leftrightarrow 
\nu_{\mu(\tau)}$ channels of the long baseline experiments and possibly also 
in the subdominant $\nu_e \leftrightarrow \nu_{\mu(\tau)}$ channels of the 
atmospheric neutrino experiments in the most plausible case of the normal 
neutrino mass hierarchy \cite{f3}. In the case of the inverted mass 
hierarchy, the same is true for the SMA solution. If the VO is established
as the true solution, we predict no observable $\nu_e \leftrightarrow 
\nu_{\mu(\tau)}$ oscillations in long baseline experiments for either of  
the mass hierarchies. 

Conversely, a measurement of $U_{e3}$ in atmospheric or long baseline 
accelerator or reactor neutrino experiments would help discriminate between 
possible oscillation solutions of the solar neutrino problem. At present, the 
situation with solar neutrinos is rather unclear: different pieces of data 
(total rates, recoil electron spectrum and day-night effect in 
Super-Kamiokande) favour different oscillation solutions, and global fits
of all the data are of comparable quality for all the solutions except LOW 
\cite{f1}. The data from the long baseline experiments could help 
clear the situation up. In particular, a positive signal of $\nu_e
\leftrightarrow \nu_{\mu(\tau)}$ oscillations would disfavour the VO 
solution of the solar neutrino problem.  

Combined data of the solar neutrino and future long baseline experiments
may provide information on the neutrino mass hierarchy even 
in the absence of any data on matter effects in the long baseline experiments.
A positive signal of $\nu_e \leftrightarrow 
\nu_{\mu(\tau)}$ oscillations along with the established LMA solution of
the solar neutrino problem would favour the normal hierarchy; 
if, however, the future solar data prefer the SMA solution, that positive 
signal would point towards the inverted neutrino mass hierarchy.

Finally, if the neutrino mass hierarchy is established through the matter
effects in the long baseline experiments, combined data of solar and long 
baseline experiments could allow one to check our assumption of no fine 
tuning between the elements of the neutrino mass matrix. 

\vglue 0.2truecm
The authors are grateful to A.Yu. Smirnov for useful discussions and to 
E. Lisi for useful correspondence. 
This work was supported in part by the TMR network grant ERBFMRX-CT960090
of the European Union. The work of E.A. was supported by Funda\c{c}\~ao
para a Ci\^encia e a Tecnologia through the grant PRAXIS XXI/BCC/16414/98.


\begin{references}
\bibitem{f2} One could also have the other solutions of the solar
neutrino problem in this case provided that the parameters $\varepsilon$ and 
$\varepsilon'$ are not small and satisfy $\varepsilon' \simeq \varepsilon$ 
\cite{Akh}. However, this is exactly the kind of the relationships between 
the matrix elements of $m_L$ that we exclude from our consideration.
\end{references}
\end{document}